\documentclass[prb,preprint,amsmath,amssymb,superscriptaddress,longbibliography]{revtex4-2}
\usepackage{graphicx}
\usepackage{hyperref}
\usepackage{xcolor}
\usepackage{bm}

\hypersetup{ pdftitle = {Resistivity in Dilute Cu-3d Alloys Governed by 
Disorder-Induced Band Broadening}, pdfauthor = {Kenji Yamaguchi}, 
pdfsubject = {}, pdfkeywords = {}, pdfproducer = {}, pdfcreator = {} }

\begin{document}

\title{Resistivity in Dilute Cu--3\textit{d} Alloys Governed by Disorder-Induced Band Broadening}

\author{Kenji Yamaguchi}
\affiliation{Innovation Center, Mitsubishi Materials Corporation, Naka, Ibaraki, Japan}

\date{\today}

\begin{abstract}
The mechanism governing the resistivity in dilute 
Cu--3\textit{d} transition-metal alloys at ambient 
temperature---the regime relevant to most practical 
applications and distinct from the low-temperature, 
Kondo-screened regime addressed by earlier theoretical 
work---is investigated using first-principles 
calculations based on the Korringa--Kohn--Rostoker 
coherent potential approximation combined with the 
Kubo--Greenwood formalism. 
The paramagnetic state is described within the 
disordered local moment (DLM) framework, corresponding 
to a local-moment paramagnet rather than a Pauli 
paramagnet.
We show that the experimentally observed resistivity trends are reproduced only within the DLM description, 
while nonmagnetic and ferromagnetic states fail to capture the correct element dependence.
Contrary to conventional interpretations based on the density of states at the Fermi level, 
the resistivity exhibits a strong correlation with the full width at half maximum (FWHM) of the Bloch spectral function (BSF) on the Fermi surface.
This correlation reflects the disorder-induced lifetime broadening of electronic states, directly related to the scattering rate that governs electrical resistivity.
A common power-law scaling between resistivity and BSF broadening is identified across different magnetic states.
These results demonstrate that the resistivity is governed by disorder-induced band broadening in momentum space rather than by local density-of-states effects, 
providing a unified microscopic interpretation of the breakdown of Linde's rule in Cu-based alloys.
\end{abstract}

\maketitle

\section{Introduction}
\label{sec:intro}

Copper-based alloys are widely used in industrial applications owing to their excellent electrical conductivity 
combined with tunable mechanical properties~\cite{Davis2001}. 
Among them, dilute solid-solution alloys are of particular importance, as their transport properties can be systematically 
controlled by alloying while maintaining relatively simple microstructures.

Beyond the design of new alloys, the demand for 
high-performance copper conductors is rapidly increasing 
in emerging applications such as electric vehicles and 
data centers, where minimizing electrical losses is 
essential both for energy efficiency and for the transition 
toward a sustainable society. In this context, 
a quantitative understanding of how trace impurities 
affect the conductivity of high-purity copper has become 
a pressing scientific and industrial need.

This need is further amplified by the growing importance 
of copper recycling. Establishing a circular economy for 
copper requires extensive recycling of end-of-life scrap, 
yet the growing complexity of scrap streams inevitably 
introduces tramp elements---unintended impurities that 
are difficult to remove by conventional metallurgical 
processes~\cite{Loibl2021}. 
Because even trace amounts of such impurities can 
significantly degrade electrical conductivity, 
a fundamental understanding of the mechanisms by which 
solute elements increase resistivity is essential not 
only for forward alloy design but also for assessing 
and ensuring the performance and reliability of recycled 
copper products.
This challenge lies at the heart of our motivation for revisiting the fundamental physics of impurity scattering in copper.

The electrical resistivity of such alloys has long been described by empirical relations, 
including Matthiessen's rule~\cite{Matthiessen1864}, the Nordheim rule~\cite{Nordheim1931}, and Linde's rule~\cite{Norbury1921,Linde1931}.
In particular, Linde's rule states that the impurity-induced increase in residual resistivity is proportional to the square of 
the valence difference between host and solute atoms.
This relation has been widely used as a guideline for alloy design in nonmagnetic dilute alloys.

Cu-based solid-solution alloys have been extensively studied from both industrial and scientific viewpoints, 
particularly for applications requiring a balance between high strength and high electrical conductivity~\cite{Davis2001,Maki2013,Ito2014}.
Experimental studies on Cu--\textit{X} alloys ($X =$ transition metals) have revealed that the resistivity 
increase shows strong element dependence and cannot be explained solely by the valence difference~\cite{Davis2001,Linde1968,Komatsu2002,Yamaguchi2023}.
In particular, for 3\textit{d} transition-metal solutes such as Cr, Mn, Fe, and Co, 
it has been consistently observed that the magnitude and trend of resistivity increase deviate significantly 
from the predictions of Linde's rule~\cite{Davis2001,Linde1968,Komatsu2002}.
These results clearly indicate that additional scattering mechanisms beyond simple charge (valence) mismatch must be considered.

From a theoretical perspective, the electronic transport properties of disordered alloys can be described within the framework of 
the Korringa--Kohn--Rostoker coherent potential approximation (KKR-CPA)~\cite{Korringa1947,Kohn1954,Soven1967,Akai1982} 
combined with the Kubo--Greenwood formalism \cite{Tulip2008,Kubo1957,Greenwood1958}.
Pioneering \textit{ab initio} calculations of residual 
resistivities for dilute Cu-based alloys with $3d$ 
transition-metal solutes were carried out by Mertig 
\textit{et al.}~\cite{Mertig1982,Mertig1999} based on the 
KKR Green's-function method combined with the Boltzmann 
equation, establishing a quantitative link between the 
impurity scattering potential and the observed resistivity 
trends across the $3d$ series. Their calculated trends were 
benchmarked against the $T \to 0$ residual resistivity 
measured at temperatures well below the Kondo temperature $T_\mathrm{K}$, 
where the local impurity moments are screened by conduction 
electrons~\cite{Kondo1964,Daybell1968,Zhu2024} and the 
system effectively behaves as a nonmagnetic scatterer. 
By contrast, the present study focuses on the resistivity 
at ambient temperature ($T \gg T_\mathrm{K}$ for most solute 
elements), where the local moments are thermally disordered 
and the paramagnetic state must be described within the disordered local moment (DLM)  
framework~\cite{Oguchi1983,Gyorffy1985}. Recent first-principles 
calculations have demonstrated that the experimentally 
observed room-temperature resistivity trends in Au- and 
Cu-based dilute alloys can be reproduced when the paramagnetic 
state is described within this DLM 
picture~\cite{Gomi2018,Yamaguchi2023}.
The importance of an explicit treatment of the paramagnetic 
state is further illustrated by a recent high-throughput 
screening of Cu alloys based on the same KKR-CPA 
methodology~\cite{Wang2022}, 
which reported an element-dependent resistivity trend for 
the $3d$ series that is inconsistent with experiment.
No specification of the magnetic configuration was provided 
in Ref.~\onlinecite{Wang2022}; as shown in the Supplemental 
Material, the reported values do not correspond to any single 
magnetic state but instead follow the ferromagnetic (FM) trend for some 
elements and the nonmagnetic (NM) trend for others, suggesting that the 
magnetic configuration was not explicitly controlled in the 
self-consistent calculations. This underscores that a 
deliberate treatment of the paramagnetic state is a 
prerequisite for a correct description of impurity scattering 
in these systems.
While the DLM description thus provides a successful 
phenomenological framework for reproducing the 
room-temperature resistivity trends, the microscopic origin 
of the element dependence remains far from fully understood. 
A widely used interpretation is based on the formation of 
virtual bound states (VBS), in which impurity \textit{d} 
states generate peaks in the density of states near the 
Fermi level, thereby enhancing scattering~\cite{Friedel1956}. 
While this picture can qualitatively account for certain 
trends, it is inherently local, focusing primarily on the 
density of states at the Fermi level. As a consequence, 
it fails to explain several key features: (i) the strong 
dependence on magnetic state (DLM, FM, and NM), 
(ii) the inconsistencies observed between different 
computational treatments, and (iii) the systematic breakdown 
of Linde's rule across the $3d$ series.
These limitations suggest that a more complete description of electronic scattering in disordered alloys must explicitly incorporate 
the momentum- and energy-resolved structure of the electronic states.
Within the KKR-CPA formalism, such information is naturally contained in the Bloch spectral function (BSF)~\cite{Faulkner1980}, 
which describes the spectral weight of electronic states in the presence of disorder.
In particular, the finite width of the BSF reflects the lifetime broadening induced by alloy disorder and spin disorder, 
providing a direct measure of the scattering strength.

In this work, we investigate the mechanism of resistivity increase in dilute Cu--3\textit{d} alloys from this perspective.
We perform systematic first-principles calculations for DLM,  FM, and NM states, 
and examine how the electronic scattering is reflected in both local quantities (density of states) and momentum-resolved quantities (BSF).
By directly comparing these different magnetic states within a unified framework, we identify the key descriptor governing the resistivity increase.
We demonstrate that the resistivity is not controlled solely by the local density of states at the Fermi level, 
but rather by the disorder-induced band broadening in momentum space.
This finding provides a unified microscopic interpretation of the resistivity behavior in Cu-based alloys and offers new insight 
into the breakdown of Linde's rule in systems containing magnetic impurities.

\section{Method}
\label{sec:method}

All calculations were performed using the Munich spin-polarized
relativistic Korringa--Kohn--Rostoker (SPR-KKR) code, version
7.7.3~\cite{SPRKKR,Ebert2011} within the framework of density functional theory~\cite{Hohenberg1964,Kohn1965}.
The exchange-correlation functional was treated within the generalized
gradient approximation of Perdew, Burke, and Ernzerhof
(PBE)~\cite{Perdew1996}.
The angular-momentum expansion of the basis was truncated at
$\ell_{\max}=3$, thereby including $f$-electron contributions, and the
full-potential mode was employed.
Brillouin-zone integrations for the self-consistent-field (SCF) potential
and total-energy calculations used 2500 $k$-points, while 10$^{5}$
$k$-points were used for the conductivity calculations.
The conductivity was confirmed to be converged with respect 
to the $k$-mesh density at this level.
The SCF cycle was converged until the total energy changed by less than
$10^{-6}$~eV/atom.
The lattice constants (3.6342$\text{\AA}$) and Debye temperatures used in the conductivity
calculations were taken from Ref.~\onlinecite{Yamaguchi2023}.

In all calculations, the host element Cu was set to 99~at.\% and the
solute element $X$ ($X$ = Ti, V, Cr, Mn, Fe, Co, Ni, Cu) to 1~at.\%.
Three magnetic configurations were considered for the electronic-structure
calculations:
(i) the DLM state,
(ii) an FM state in which a finite initial magnetic moment 
was imposed on the solute atom $X$, and
(iii) an NM state in which the initial moment was set to 
zero and confirmed to remain zero upon convergence.
For each of the three converged SCF potentials, the electrical conductivity
was evaluated using the Kubo--Greenwood formula with the vertex correction~\cite{Tulip2008}.
In this way, two distinct theoretical models for the paramagnetic
state---DLM and NM---were compared.

\section{Results and Discussion}
\label{sec:results}

\subsection{Magnetic ground state and Curie temperature}
\label{sec:curie}

In the SCF calculations for $X$ = Ti, Ni, and Cu the magnetic moment
converged to numerically negligible values regardless of the initial
conditions in both the DLM and FM configurations.
For $X$ = V through Co, the FM state was found to be more stable than the
NM state in every case, indicating that the zero-moment state is a
metastable configuration.

\begin{figure}[tb]
  \centering
  \includegraphics[width=\columnwidth]{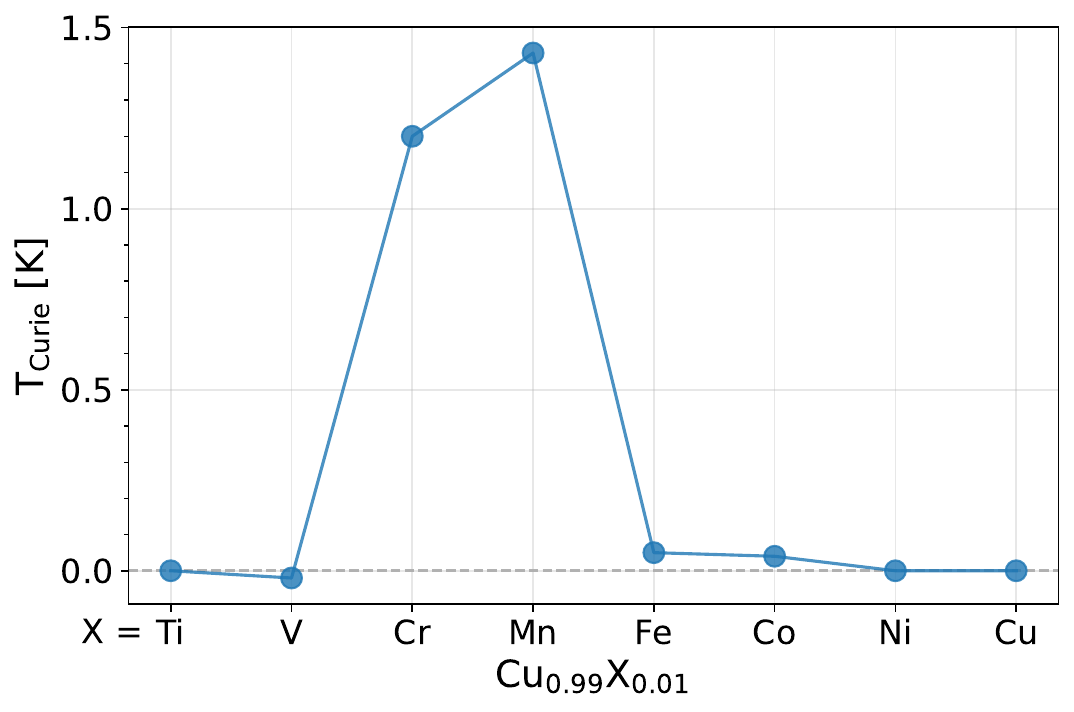}
  \caption{Calculated Curie temperatures for Cu$_{0.99}X_{0.01}$ alloys
    ($X$ = Ti--Cu) estimated from Eq.~\eqref{eq:Tc} within the mean-field
    approximation.  The dashed line indicates $T_{\mathrm{C}}=0$.}
  \label{fig:curie}
\end{figure}

The Curie temperature was estimated within the mean-field
approximation~\cite{Sato2003}:
\begin{equation}
  T_{\mathrm{C}}
  = \frac{2}{3 k_{\mathrm{B}}}
    \frac{E_{\mathrm{DLM}} - E_{\mathrm{FM}}}{x},
  \label{eq:Tc}
\end{equation}
where $k_{\mathrm{B}}$ is the Boltzmann constant, $x = 0.01$ is the
solute concentration, and $E_{\mathrm{DLM}}$ and $E_{\mathrm{FM}}$ denote
the total energies in the DLM and FM states, respectively.
The total energies were optimized by evaluating several lattice-constant
values around the value fixed in the conductivity calculations.
The results are shown in Fig.~\ref{fig:curie}.

The estimated Curie temperatures do not exceed 1.5~K for any solute
element, confirming that at ambient temperature (300~K) the system is in
the paramagnetic state described by the Curie--Weiss law within the DLM
picture.
For $X$ = V, a negative Curie temperature was obtained, suggesting that an
antiferromagnetic or other non-FM magnetic structure is more stable than
the FM state.
Nevertheless, given the small energy difference between the magnetic
configurations, the DLM description remains valid at room temperature, and
the details of the magnetic structure were not pursued further.

It is worth noting the relationship between the present DLM description and the Kondo effect~\cite{Kondo1964}. 
Dilute $3d$ impurities in Cu constitute a classic Kondo system~\cite{Daybell1968,Zhu2024}: 
below the Kondo temperature $T_\mathrm{K}$, the impurity local moment is screened by conduction electrons, 
forming a many-body singlet state. 
For $3d$ impurities in Cu, $T_\mathrm{K}$ spans several orders of magnitude—from 
$\sim10^{-2}$~K for Mn to $\sim10^{3}$~K for Ti~\cite{Daybell1968,Zhu2024}. 
For elements with $T_\mathrm{K} \ll 300$~K (e.g., Mn, Fe), the local moments remain unscreened 
at ambient temperature, which is precisely the situation described by the DLM framework. 
For elements with $T_\mathrm{K} \gg 300$~K (e.g., Ti, Ni), the moments are fully screened 
and effectively absent—consistent with our SCF finding that these solutes converge to zero magnetic moment. 
In either case, the DLM treatment at 300~K introduces no inconsistency with the known Kondo physics of dilute Cu alloys.
This temperature-regime distinction also reconciles the 
present work with the pioneering low-temperature calculations 
of Mertig \textit{et al.}~\cite{Mertig1982,Mertig1999}: 
as shown in the Supplemental Material, their results 
quantitatively agree with our NM calculations, consistent 
with the Kondo-screened, effectively nonmagnetic regime 
probed by the $T \to 0$ residual-resistivity measurements 
they used for benchmarking.

\subsection{Resistivity: comparison among DLM, FM, and NM states}
\label{sec:resistivity}

\begin{figure}[tb]
  \centering
  \includegraphics[width=\columnwidth]{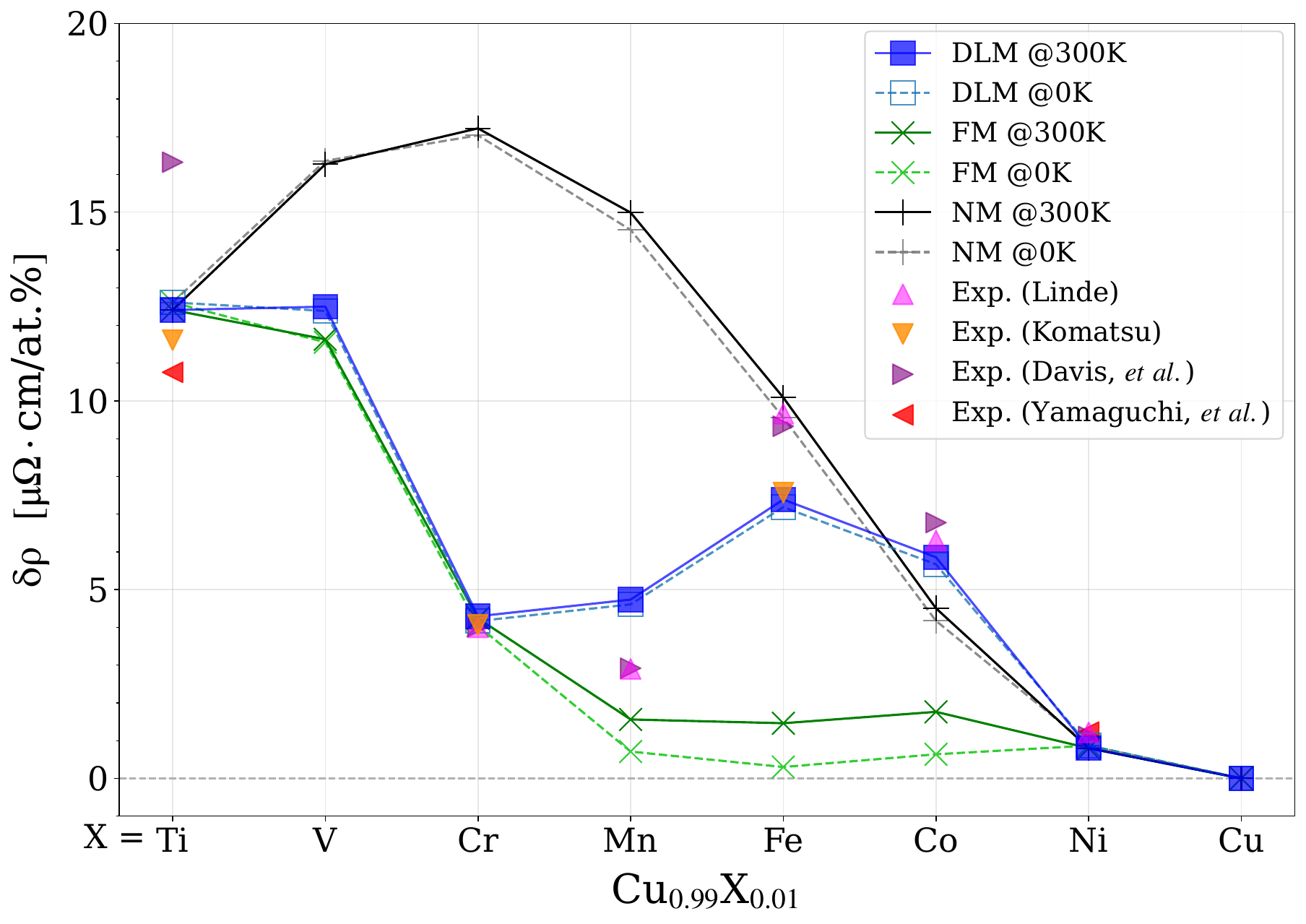}
  \caption{Calculated and experimental electrical resistivity increase
    $\delta\rho$ per 1~at.\% of solute $X$ relative to pure Cu, obtained
    from first-principles KKR-CPA calculations in the DLM, FM, and NM
    states at 0~K and 300~K, together with experimental
    data~\cite{Davis2001,Linde1968,Komatsu2002,Yamaguchi2023}.}
  \label{fig:resistivity}
\end{figure}

We now compare the calculated resistivities in the DLM, FM, 
and NM states at $T = 0$~K and $T = 300$~K with 
room-temperature experimental data 
(Fig.~\ref{fig:resistivity}). As outlined in 
Sec.~\ref{sec:curie}, the DLM and NM treatments respectively 
describe the high-temperature ($T \gg T_\mathrm{K}$) and the 
low-temperature ($T \ll T_\mathrm{K}$) regimes of the same 
dilute alloy. A detailed comparison of our NM results with 
the earlier Mertig calculations~\cite{Mertig1999} and the 
$T \to 0$ experimental data~\cite{Landolt1982} is presented 
in the Supplemental Material.

As can be seen from Fig.~\ref{fig:resistivity}, the experimental results
from multiple reports are well reproduced by the DLM calculations.
Matthiessen's rule~\cite{Matthiessen1864} is closely obeyed: the temperature
dependence is negligible, and the element dependence of the resistivity can
be explained by the residual resistivity at 0~K arising from the
randomness associated with elemental substitution and spin-orientation
disorder.
The NM state also obeys Matthiessen's rule well, as evident from
Fig.~\ref{fig:resistivity}.
The largest difference between the 0~K and 300~K results was found for the
FM state of the magnetic elements; however, the deviation from experiment
is substantial in all cases, and no qualitative difference in the
element-dependent trend was observed.
Since Matthiessen's rule is approximately satisfied in all three states
(Fig.~\ref{fig:resistivity}), the following discussion of the resistivity
mechanism is based on the $T=0$~K results.

\subsection{Partial density of states at the Fermi level}
\label{sec:pdos}

Regarding the resistivity increase per atom\% upon $3d$-element addition to
Au, Gomi and Yoshino~\cite{Gomi2018} explained the higher resistivity in
the NM state compared with the DLM state in terms of the virtual bound
state (VBS) mechanism~\cite{Friedel1956}: the DOS peak of the
solute-derived $d$ states located near the Fermi level acts as a scattering
center for conduction electrons.
In particular, for Cr the $3d$ band in the DLM state splits above and
below the Fermi level, resulting in a low partial DOS (pDOS) of Cr $3d$
states at $E_{\mathrm{F}}$, whereas in the NM state no such splitting
occurs and the pDOS at $E_{\mathrm{F}}$ is high, leading to a larger
resistivity increase.
Conversely, for Co the high-energy peak of the exchange-split $3d$ band in
the DLM state overlaps with the Fermi level, making the DLM resistivity
higher than that in the NM state.

\begin{figure}[tb]
  \centering
  \includegraphics[width=\columnwidth]{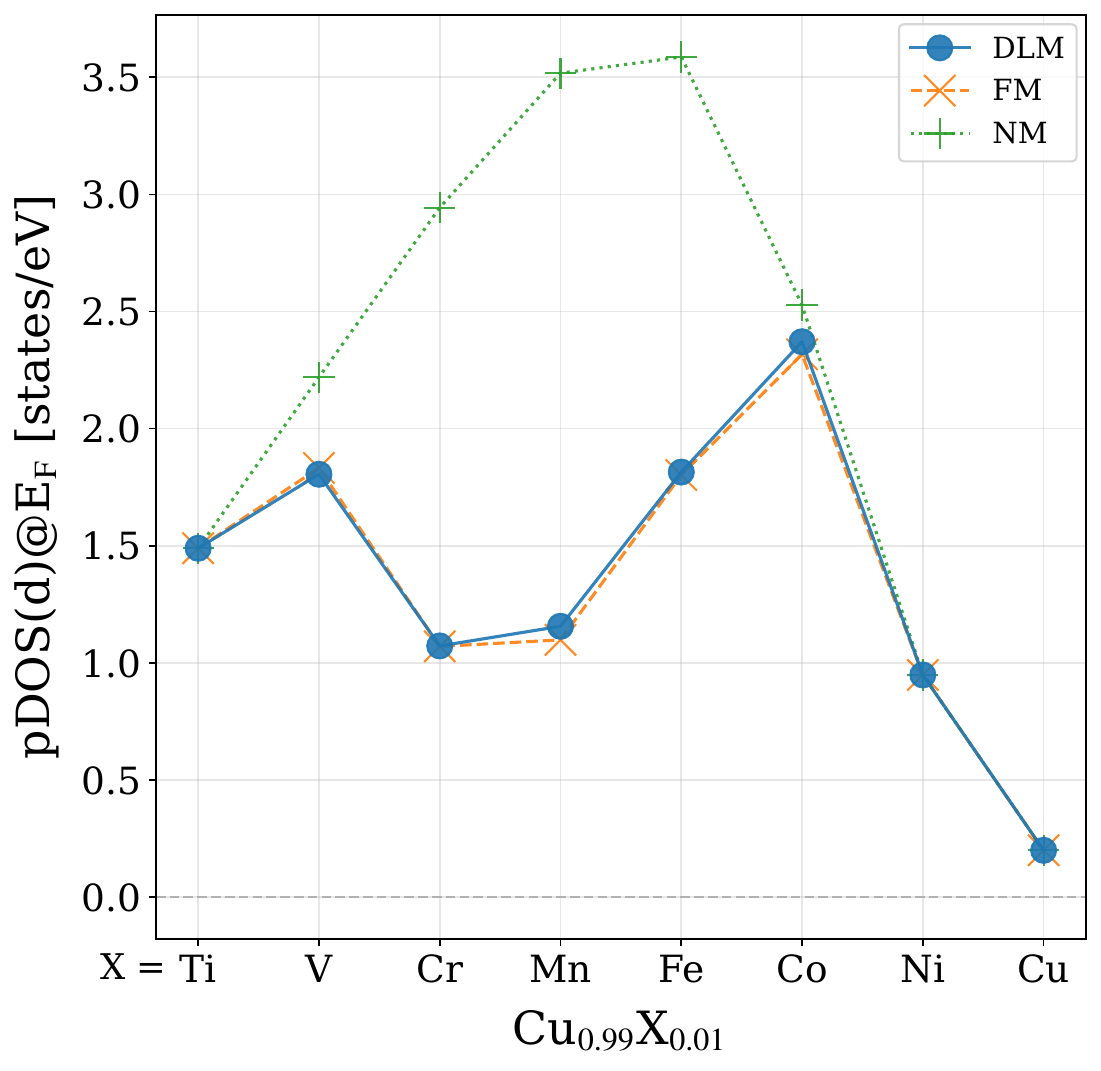}
  \caption{Partial density of states (pDOS) of the solute-element $X$
    $d$-states at the Fermi level for Cu$_{0.99}X_{0.01}$ in the DLM, FM,
    and NM states.}
  \label{fig:pdos}
\end{figure}

Figure~\ref{fig:pdos} shows the pDOS of the solute-element $d$ states at
$E_{\mathrm{F}}$ for 1~at.\% $X$ additions to Cu in the DLM, FM, and NM
states.
While the pDOS values in Fig.~\ref{fig:pdos} can explain the enhanced
resistivity increase in the NM state for $X$ = V, Cr, Mn, and Fe, the
difference between the DLM and FM states is only marginal.
Thus, a comparison of the DOS at the Fermi level based on the VBS picture
alone is insufficient to account for the resistivity difference between
these two states shown in Fig.~\ref{fig:resistivity}.

\subsection{Bloch spectral function}
\label{sec:bsf}

We therefore turn to the Bloch spectral function
(BSF)~\cite{Faulkner1980}, a quantity computed within KKR-CPA that
reflects the scattering arising from both compositional disorder due to
solute atoms and spin-orientation randomness.
In a perfect crystal the band dispersion is represented by $\delta$
functions in the spectral weight at well-defined wave vectors $\bm{k}$;
in a disordered alloy, however, $\bm{k}$ is no longer a good quantum
number and the spectral weight broadens into a Lorentzian-like profile
whose width reflects the degree of disorder-induced scattering.

\begin{figure}[tb]
  \centering
  \includegraphics[width=\columnwidth]{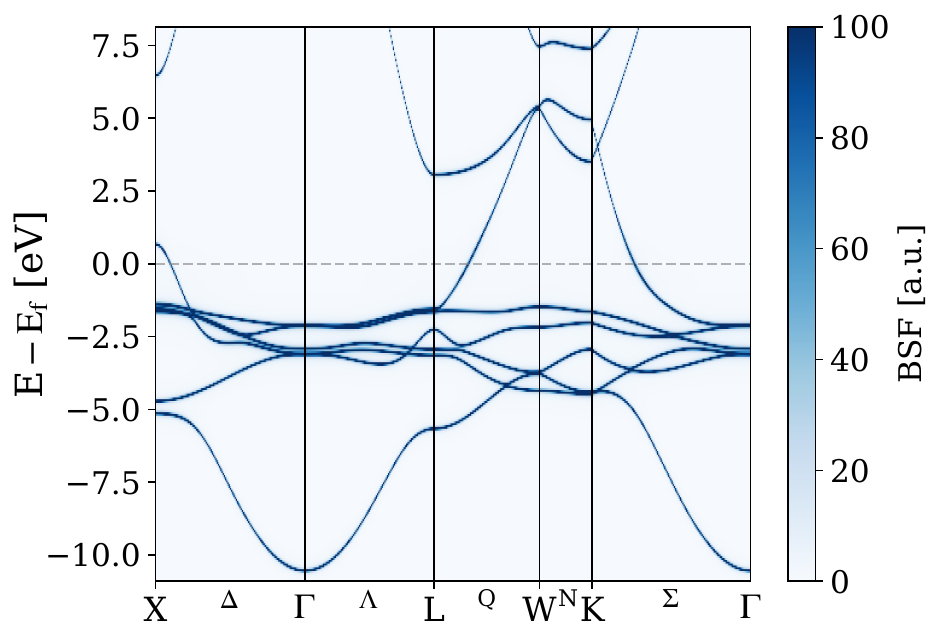}
  \caption{Bloch spectral function (BSF) dispersion of
  Cu$_{0.99}$Fe$_{0.01}$ in the DLM state along high-symmetry
  directions. The color scale represents the BSF intensity in 
  arbitrary units.}
  \label{fig:bsf_dispersion}
\end{figure}

Figure~\ref{fig:bsf_dispersion} shows the BSF of Cu$_{0.99}$Fe$_{0.01}$
in the DLM state.
The overall BSF dispersion closely resembles the band structure of pure fcc Cu, with
Fermi-surface crossings on the $\Delta$, $Q$, and $\Sigma$ lines.
Regions where the BSF intensity is weaker (lighter shading) correspond to
stronger scattering due to solute dissolution and spin-orientation
disorder.
Because the solute concentration is only 1~at.\%, the BSF spectra in the
DLM, FM, and NM states are nearly indistinguishable in shape from the
pure-Cu band structure, differing only in the degree of broadening.

\subsection{BSF on the Fermi surface}
\label{sec:bsf_fermi}

\begin{figure}[tb]
  \centering
  \includegraphics[width=\columnwidth]{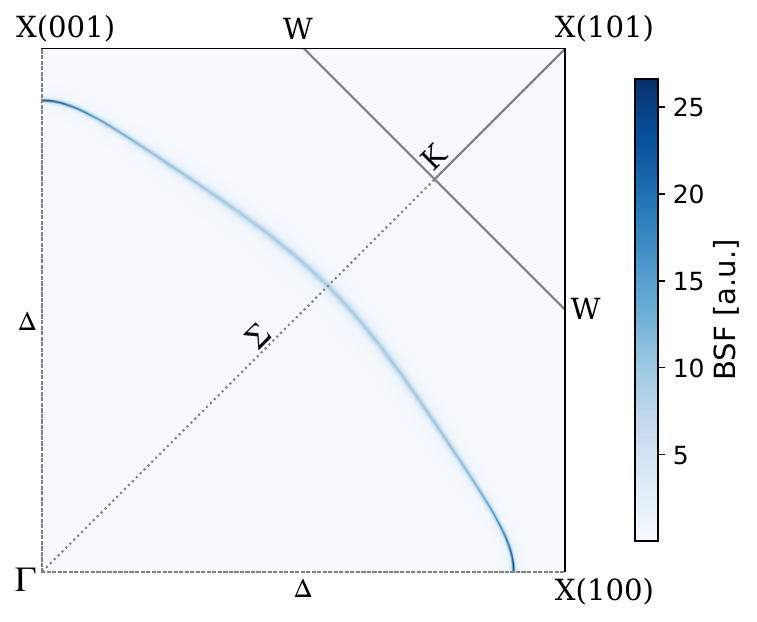}
  \caption{BSF at the Fermi energy of Cu$_{0.99}$Fe$_{0.01}$ 
  in the DLM state, plotted on the plane defined by 
  $\Gamma$--X$(100)$ and $\Gamma$--X$(001)$. 
  The color scale represents the BSF intensity.}
  \label{fig:bsf_fermi}
\end{figure}

To examine the Fermi-surface broadening in more detail, we plot the BSF at
the Fermi energy on the plane spanned by the two orthogonal axes connecting
$\Gamma$ to X$(100)$ and $\Gamma$ to X$(001)$.
Figure~\ref{fig:bsf_fermi} shows this quantity for Cu$_{0.99}$Fe$_{0.01}$
in the DLM state as a representative example.
The Fermi-surface cross section in this plane forms a single arc starting
from a point $(k_{d},0,0)$ on the $\Delta$ line, crossing the $\Sigma$
line, and terminating at a point $(0,0,k_{d})$ on the $\Delta$ line.

\begin{figure}[tb]
  \centering
  \includegraphics[width=\columnwidth]{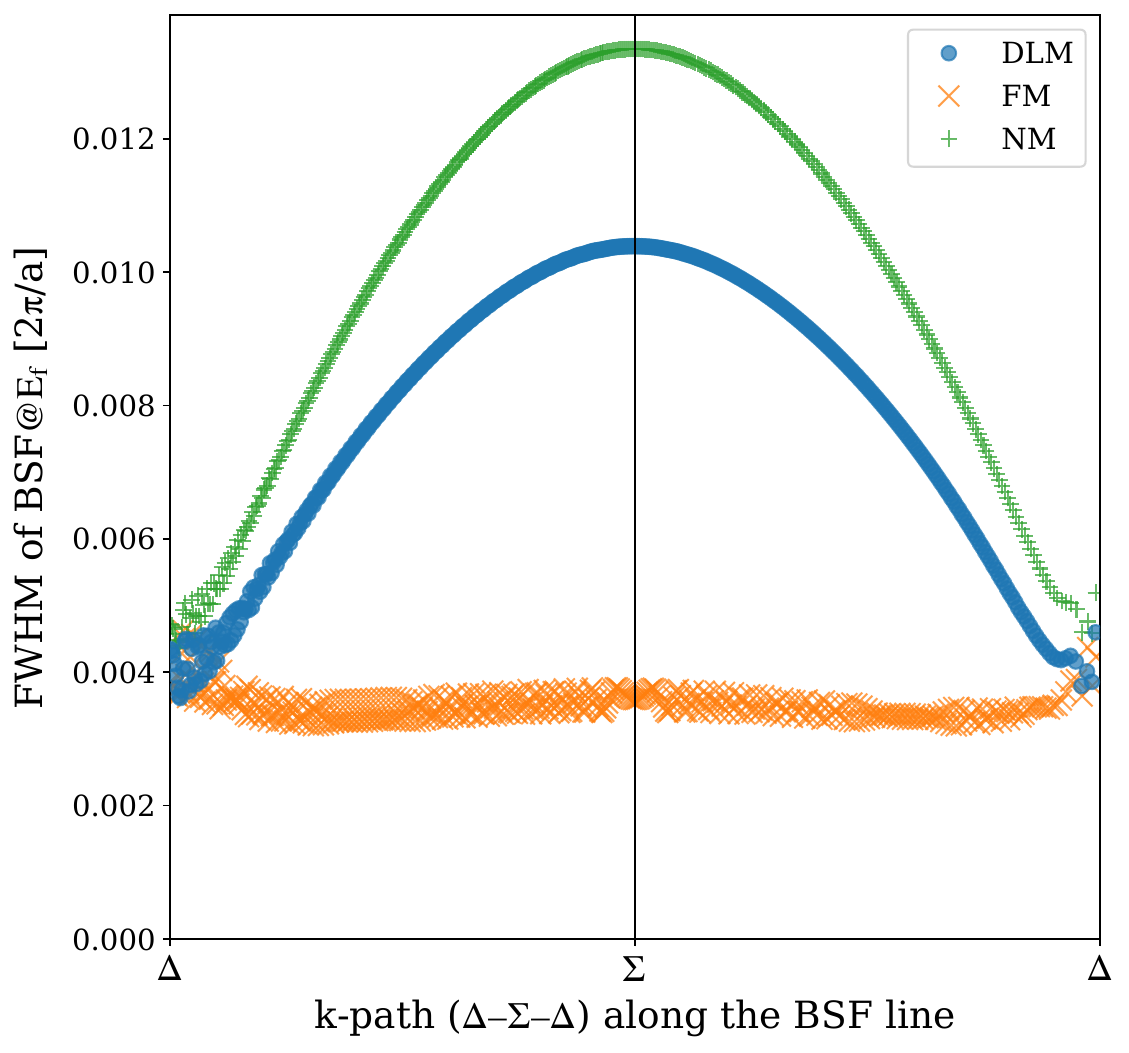}
  \caption{FWHM of the BSF along the Fermi-surface arc
  in the $\Gamma$--X$(100)$--X$(001)$ plane for
  Cu$_{0.99}$Fe$_{0.01}$ in the DLM, FM, and NM states.
  The vertical line marks the $\Sigma$ point.}
  \label{fig:fwhm_line}
\end{figure}

At each point along this Fermi-surface arc, the BSF profile in the
direction perpendicular to the local tangent (and lying within the plane)
was fitted to a Lorentzian function to extract the full width at half
maximum (FWHM).
Figure~\ref{fig:fwhm_line} compares the FWHM along the BSF line for
Cu$_{0.99}$Fe$_{0.01}$ in the DLM, FM, and NM states.
The difference among the three magnetic states is most pronounced on the
$\Sigma$ line.

\begin{figure}[tb]
  \centering
  \includegraphics[width=\columnwidth]{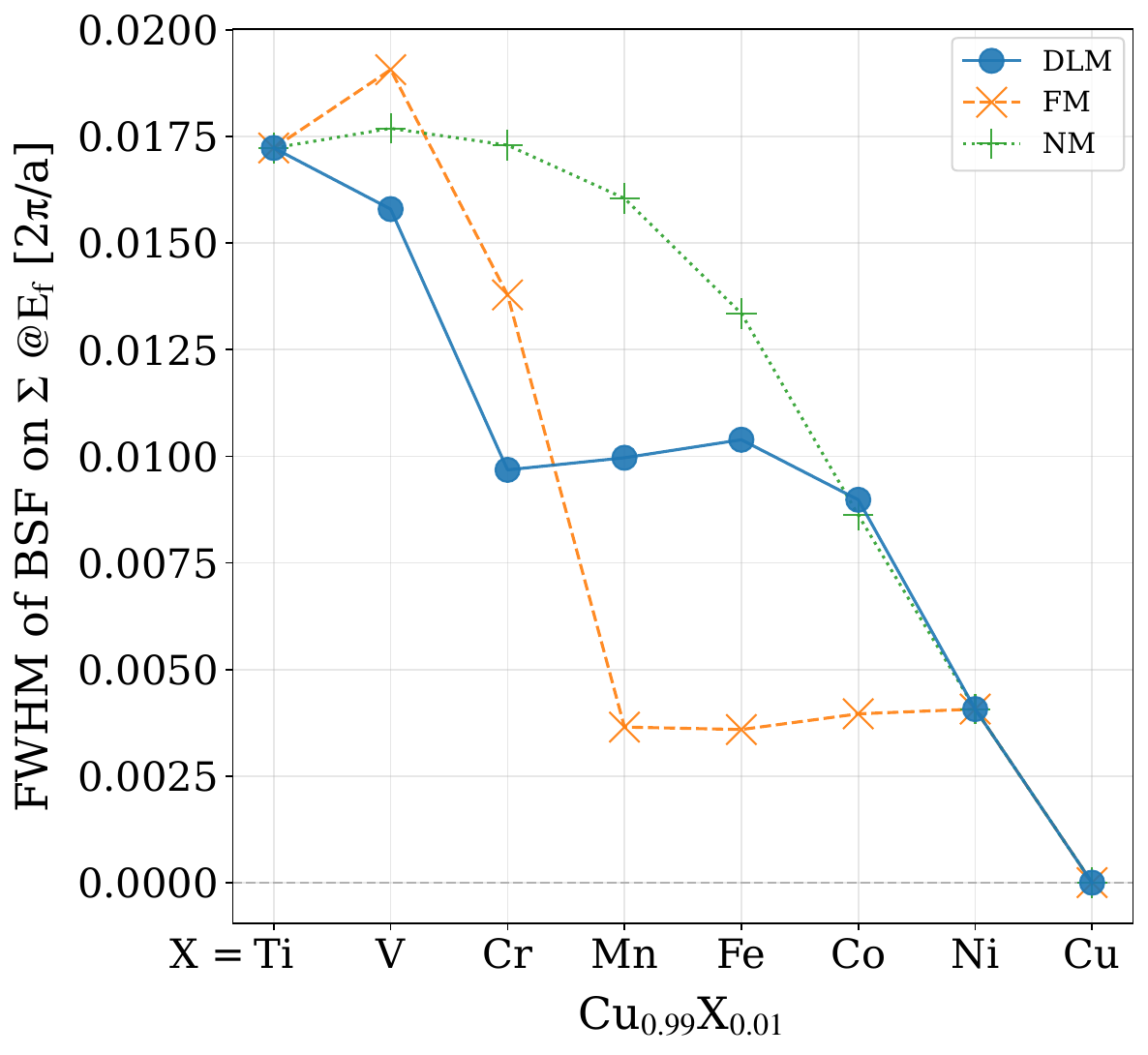}
  \caption{FWHM of the BSF at the Fermi-surface crossing 
  on the $\Sigma$ line for Cu$_{0.99}X_{0.01}$ alloys 
  in the DLM, FM, and NM states.}
  \label{fig:fwhm_sigma}
\end{figure}

Figure~\ref{fig:fwhm_sigma} shows the FWHM of the BSF on the $\Sigma$-line
crossing of the Fermi surface for all solute elements $X$ in the DLM, FM,
and NM states.
For Co, Fe, and Mn, where the DLM resistivity is lower than the NM
resistivity, the same ordering (DLM $>$ FM) is reflected in the FWHM of the BSF.
The element- and state-resolved FWHM profiles along the full Fermi-surface arc are shown in Figs. S1--S7 
of the Supplemental Material, confirming that the discrimination arises predominantly on the $\Sigma$ line.
We note that the FWHM evaluated on the $\Delta$ line 
shows weaker discrimination among the three magnetic 
states than that on the $\Sigma$ line: although the 
absolute FWHM varies across solute elements, the 
ordering among DLM, FM, and NM states is less 
systematic. 
In contrast, the $\Sigma$ direction, corresponding to the nearest-neighbor $\langle 110 \rangle$ direction in real space, 
exhibits clear differentiation. 
This is physically consistent with the fact that scattering is most strongly affected along the direction connecting 
nearest-neighbor sites, where the potential mismatch between host and solute atoms is most directly probed.

\subsection{Correlation between resistivity and BSF broadening}
\label{sec:correlation}

We now examine which quantity---the pDOS of solute $d$ states at
$E_{\mathrm{F}}$ (Fig.~\ref{fig:pdos}) or the FWHM of the BSF on the $\Sigma$
line (Fig.~\ref{fig:fwhm_sigma})---better captures the solute-element
dependence of the resistivity increase calculated via the Kubo--Greenwood
formula at $T=0$~K (Fig.~\ref{fig:resistivity}).

\begin{figure}[tb]
  \centering
  \includegraphics[width=\columnwidth]{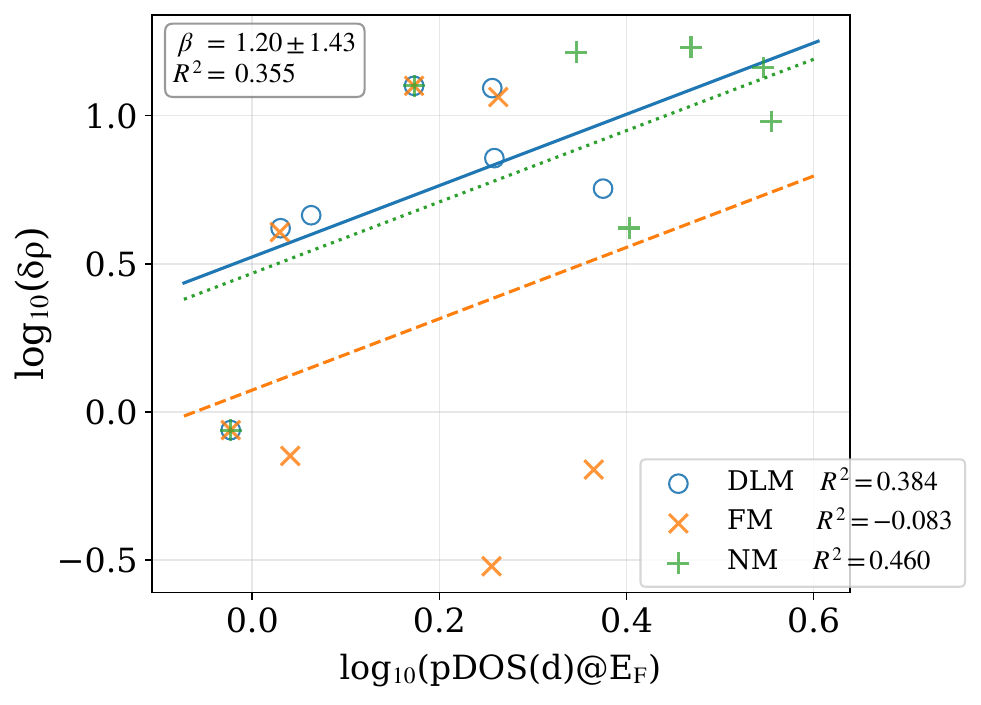}
  \caption{Log--log plot of the resistivity increase $\delta\rho$ versus
    the pDOS of the solute $d$ states at $E_{\mathrm{F}}$ for
    Cu$_{0.99}X_{0.01}$ in the DLM, FM, and NM states.
    Lines indicate fits to a common power-law model; 
    the poor $R^2$ and large slope uncertainty confirm that 
    the pDOS at $E_\mathrm{F}$ does not serve as a reliable descriptor of the resistivity.}
  \label{fig:rho_vs_pdos}
\end{figure}

Figure~\ref{fig:rho_vs_pdos} presents a log--log plot of the resistivity
increase $\delta\rho$ versus the pDOS of the solute $d$ states at
$E_{\mathrm{F}}$.
No clear correlation is observed: the pDOS at the Fermi level does not
serve as a reliable descriptor of the resistivity increase across the
$3d$ series.

\begin{figure}[tb]
  \centering
  \includegraphics[width=\columnwidth]{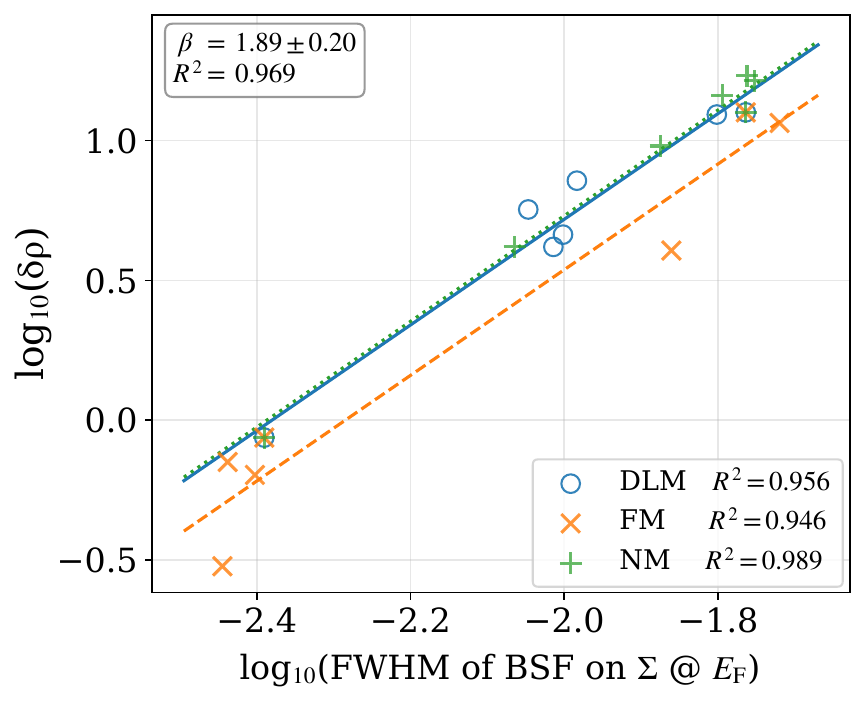}
  \caption{Correlation between the resistivity increase 
  $\delta\rho$ and the FWHM of the BSF on the $\Sigma$ line 
  at $E_\mathrm{F}$ for Cu$_{0.99}X_{0.01}$ in the 
  DLM, FM, and NM states. Lines are fits with a common 
  power-law exponent ($\beta \approx 1.89$); 
  the high individual $R^2$ values (legend) confirm that 
  the common scaling holds well across all three magnetic 
  configurations.}
  \label{fig:rho_vs_fwhm}
\end{figure}

In contrast, Fig.~\ref{fig:rho_vs_fwhm} shows that the resistivity
increase and the FWHM of the BSF on the $\Sigma$ line exhibit an excellent
correlation.
A common power-law exponent of 1.89 was obtained for all three magnetic
states (DLM, FM, and NM) by fitting with a shared slope.
The non-unity exponent ($\approx 1.89$) deserves comment. 
Within a simple relaxation-time picture, the resistivity would scale linearly 
with the scattering rate $\it{\Gamma}$, i.e., $\delta\rho \propto \it{\Gamma}^1$. 
However, the Kubo--Greenwood conductivity is not a single-channel quantity; 
it involves a Fermi-surface integral weighted by the group velocities 
and transition-matrix elements at every $\mathbf{k}$-point. 
The FWHM of the BSF evaluated at a single representative point on the $\Sigma$ line 
therefore does not capture the full integration measure: 
the Fermi-surface curvature, the anisotropy of the velocity field, 
and the $\mathbf{k}$-dependent variation of the spectral broadening 
all contribute to the mapping between the local point on $\Sigma$-line  
and the integrated resistivity. 
The deviation of the exponent from unity is a direct manifestation of these geometric and kinematic factors. 
Crucially, the fact that a common exponent is obtained for the DLM, FM, and NM states indicates that 
these weighting factors are governed by the topology of the host Cu Fermi surface 
which is virtually identical across the three magnetic configurations at 1 at.\% solute 
rather than by the magnetic state of the impurity. 
The power-law exponent thus encodes the intrinsic geometry of the Fermi surface 
through which the disorder-induced scattering is filtered into a macroscopic resistivity.

These results demonstrate that the resistivity in Cu 
solid-solution alloys is not governed by the local density 
of states of the solute $d$ states at the Fermi level, but 
rather by the FWHM of the BSF, which corresponds to the 
imaginary part of the electronic self-energy and is 
therefore inversely proportional to the quasiparticle 
lifetime. In contrast to the pDOS, which characterizes the 
\textit{availability} of electronic states at $E_\mathrm{F}$, 
the FWHM of the BSF captures the \textit{decoherence} of 
Bloch states induced by disorder. The present results 
indicate that it is this decoherence---rather than the 
local density of scattering channels---that determines the 
resistivity, with a single underlying scattering mechanism, 
originating from the solute element, operating in the DLM, 
FM, and NM configurations alike.

It should be emphasized that the FWHM of the BSF and the Kubo--Greenwood resistivity are not trivially related, 
despite both being derived from the CPA Green's function. The Kubo--Greenwood formula involves a Fermi-surface integral 
that incorporates contributions from all $\mathbf{k}$-points, whereas the FWHM of the BSF is a local quantity in momentum space 
evaluated at a specific $\mathbf{k}$-point. Indeed, the FWHM evaluated on the $\Delta$ line 
does not yield a consistent correlation across the three magnetic states 
with the resistivity (see Supplemental Material), demonstrating that the correlation 
is specific to the $\Sigma$ direction and therefore nontrivial. 
The present analysis thus identifies the dominant scattering channel within the full Fermi-surface integral: 
it is the disorder-induced broadening along the nearest-neighbor $\langle 110 \rangle$ direction 
that governs the resistivity, rather than a uniform broadening over the entire Fermi surface.

The lower resistivity in the FM state is consistent with
the expected suppression of spin-flip scattering in the
presence of exchange splitting, although a quantitative
decomposition into spin-conserving and spin-flip channels
is beyond the scope of the present work.

\section{Conclusion}
\label{sec:conclusion}
We have investigated the mechanism of resistivity increase 
in dilute Cu--3\textit{d} transition-metal alloys at ambient 
temperature using first-principles KKR-CPA calculations 
combined with the Kubo--Greenwood formalism.

The estimated Curie temperatures for all solute elements 
fall below 1.5~K, confirming that the paramagnetic state 
at 300~K is appropriately described by the DLM framework. 
This conclusion is further supported by its consistency 
with the known Kondo physics of dilute Cu alloys. 
The experimentally observed element-dependent resistivity 
trends are quantitatively reproduced only within the DLM 
description, while NM calculations yield an inverted trend 
inconsistent with room-temperature experiment---a finding 
that also accounts for the discrepancy reported in a recent 
computational screening study (see Sec. II of the Supplemental Material).

Contrary to conventional interpretations based on the 
virtual bound state picture, the partial density of states 
of the solute \textit{d} states at the Fermi level does not 
serve as a reliable descriptor of the resistivity increase 
across the $3d$ series. Instead, the resistivity exhibits 
a strong correlation with the FWHM of the BSF on the $\Sigma$ line 
of the Fermi surface, following a common power-law scaling 
with an exponent of approximately 1.89 across the DLM, FM, 
and NM states. This common exponent reflects the intrinsic 
geometry of the host Cu Fermi surface rather than the 
magnetic configuration of the impurity. Notably, the 
correlation is specific to the $\Sigma$ direction---corresponding to the nearest-neighbor 
$\langle 110 \rangle$ direction in real space---and is 
absent on the $\Delta$ line, demonstrating that the 
relationship between the BSF broadening and the 
Kubo--Greenwood resistivity is nontrivial.

These results provide a unified microscopic interpretation 
of the breakdown of Linde's rule in Cu-based alloys 
containing magnetic impurities: the resistivity is governed 
by disorder-induced band broadening along the directions 
most sensitive to the solute-induced potential mismatch, 
rather than by local density-of-states effects. The FWHM of the BSF 
thus serves as a physically transparent and 
computationally accessible descriptor for resistivity in 
disordered metallic alloys.

Finally, we note that the quantitative comparison of 
theoretical predictions with experimental residual 
resistivities depends sensitively on the purity of the 
host metal used in the measurements. The systematic 
development of ultra-high-purity copper over recent 
decades---progressing from the 4--5N grades available 
when most of the foundational $T \to 0$ data were 
compiled to the 6--7N grades produced industrially 
today---has steadily reduced the contribution of the host 
itself to the measured residual 
resistivity~\cite{Kurosaka1990,Li1988,Mimura1997}. Continued 
advances in metal refining are therefore an essential 
counterpart to the theoretical efforts pursued here, and 
the increasing availability of well-characterized 
high-purity samples will allow ever more stringent tests 
of first-principles predictions for the electrical 
transport in dilute alloys.

The author thanks Dr. K. Maki, Dr. H. Mori, and K. Suehiro 
of the Copper \& Copper Alloy Products Division, 
Mitsubishi Materials Corporation, 
for discussions that gradually clarified what needed to be done 
and quietly motivated the present work.

Microsoft 365 Copilot was used in both its automatic and Opus modes 
between May and June 2026 to assist with English-language editing 
and translation of author-drafted text, to discuss possible directions 
of the argument, and to suggest relevant literature. 
All AI-suggested references were independently verified against the 
primary sources by the author, and all scientific ideas, analyses, 
and conclusions are the author's own. 
The author takes full responsibility for the entire content 
of this manuscript. 

\bibliographystyle{apsrev4-2}
\bibliography{refs}

\end{document}